\documentclass[lettersize,journal]{IEEEtran}
\usepackage{amsmath,amsfonts}
\usepackage{algorithmic}
\usepackage{algorithm}
\usepackage{array}
\usepackage[caption=false,font=normalsize,labelfont=sf,textfont=sf]{subfig}
\usepackage{textcomp}
\usepackage{stfloats}
\usepackage{url}
\usepackage{verbatim}
\usepackage{graphicx}
\usepackage{cite}

\usepackage[colorinlistoftodos]{todonotes}
\usepackage[normalem]{ulem}
\usepackage{soul}

\usepackage{booktabs}
\usepackage{outlines}
\usepackage{lipsum}
\usepackage{xcolor}
\usepackage[T1]{fontenc}
\usepackage{url}

\newcommand{\etal}{\textit{~et~al.}}

\begin{document}

\title{Deep Learning for Detection and Localization of B-Lines in Lung Ultrasound}
 
\author{
R.~T.~Lucassen, M.~H.~Jafari, N.~M.~Duggan, 
N.~Jowkar, A.~Mehrtash, C.~Fischetti, D.~Bernier,
K.~Prentice, E.~P.~Duhaime, M.~Jin,
P.~Abolmaesumi, F.~G.~Heslinga, M.~Veta, 
M.~A.~Duran-Mendicuti, S.~Frisken, P.~B.~Shyn, A.~J.~Golby, E.~Boyer, W.~M.~Wells, A.~J.~Goldsmith, T.~Kapur
\thanks{This work was supported by the US National Institutes of Health grant R01EB027134-S1, and the Massachusetts Life Sciences Center Bits to Bytes Award. (R.~T.~Lucassen, M.~H.~Jafari, and N.~Duggan are co-first authors.) (T.~Kapur and A.~J.~Goldsmith are  co-senior authors.)}
\thanks{R.~T.~Lucassen, M.~H.~Jafari, 
N.~Jowkar, A.~Mehrtash, M.~Jin, M.~A.~Duran-Mendicuti, S.~Frisken, P.~B.~Shyn, W.~M.~Wells, and T.~Kapur are with the Department of Radiology, Brigham and Women’s Hospital, Harvard Medical School, Boston, MA 02115 USA (e-mail: tkapur@bwh.harvard.edu).}
\thanks{R.~T.~Lucassen, F.~G.~Heslinga, and M.~Veta are with the Department of Biomedical Engineering, Eindhoven University of Technology, 5612 Eindhoven, The Netherlands.}
\thanks{N.~M.~Duggan, C.~Fischetti, D.~Bernier, A.~J.~Goldsmith, and E.~Boyer are with the Department of Emergency Medicine, Brigham and Women's Hospital, Boston, MA 02115 USA.}
\thanks{K.~Prentice, E.~P.~Duhaime, and M.~Jin are with Centaur Labs, Boston, MA 02116 USA.}
\thanks{P.~Abolmaesumi is with the Department of Electrical and Computer Engineering, The University of British Columbia, Vancouver, BC V5T 1Z4 Canada.}
\thanks{A.~J.~Golby is with the Departments of Neurosurgery and Radiology, Brigham and Women’s Hospital, Harvard Medical School, Boston, MA 02115 USA.}
}

\maketitle

\begin{abstract} 
Lung ultrasound (LUS) is an important imaging modality used by emergency physicians to assess pulmonary congestion at the patient bedside. B-line artifacts in LUS videos are key findings associated with pulmonary congestion. Not only can the interpretation of LUS be challenging for novice operators, but visual quantification of B-lines remains subject to observer variability. 
In this work, we investigate the strengths and weaknesses of multiple deep learning approaches for automated B-line detection and localization in LUS videos. We curate and publish, \textit{BEDLUS}, a new ultrasound dataset comprising 1,419 videos from 113 patients with a total of 15,755 expert-annotated B-lines. Based on this dataset, we present a benchmark of established deep learning methods applied to the task of B-line detection. To pave the way for interpretable quantification of B-lines, we propose a novel "single-point" approach to B-line localization using only the point of origin.
Our results show that (a) the area under the receiver operating characteristic curve ranges from 0.864 to 0.955 for the benchmarked detection methods, (b) within this range, the best performance is achieved by models that leverage multiple successive frames as input, and (c) the proposed single-point approach for B-line localization reaches an F$_{1}$-score of 0.65, performing on par with the inter-observer agreement. The dataset and developed methods can facilitate further biomedical research on automated interpretation of lung ultrasound with the potential to expand the clinical utility.

\end{abstract}

\section{Introduction}
\label{sec:introduction}

Lung ultrasonography (LUS) has emerged as an important bedside imaging modality to support diagnostic assessment and therapeutic management in acute care settings \cite{volpicelli2020lung, sekiguchi2015critical, pivetta2015lung, moore2011point}.
B-line artifacts in LUS are defined as hyperechoic lines that originate from the pleura and extend radially to the bottom of the screen, while moving synchronously with respiration \cite{ lichtenstein1997comet,volpicelli2012international}.
Despite the artifactual nature, B-lines play a key role in detecting and assessing the severity of pulmonary congestion in patients with disease states such as decompensated heart failure, chronic kidney disease on dialysis, viral and bacterial pneumonia, or interstitial lung disease \cite{volpicelli2012international,lichtenstein1997comet,agricola2005ultrasound,platz2017dynamic,mallamaci2010detection,noble2009ultrasound,gargani2020prognostic}. The quantity of B-lines is in most cases a biomarker for pulmonary congestion that drives treatment decisions. In clinical practice, physicians generally assign scores for disease severity based on a visual estimate of the quantity of B-lines. However, several studies have illustrated substantial inter- and intra-observer variability in B-line quantification \cite{gullett2015interobserver,pivetta2018sources, matthias2021effect}, with differences in expertise, operator dependence, and acquisition settings as possible causes. 

The COVID-19 pandemic catalyzed the development of automated methods for reproducible LUS assessment \cite{roy2020deep, born2021accelerating, frank2021integrating, mento2021deep}. Well before polymerase chain reaction (PCR) tests became widely available, lung ultrasound was heavily used for COVID-19 triage in acute care settings \cite{soldati2020there, smith2020point, duggan2020using}. B-lines helped determine which patients needed to be closely monitored, and clinical guidelines on severity assessment of COVID-19 for LUS emerged. Tools developed for computer-assisted LUS assessment can potentially improve the inter- and intra-observer agreement, reduce diagnostic error, and assist novice ultrasound users in their education and patient examinations \cite{baloescu2020automated, russell2021b}. However, a direct comparison of methods that have thus far been proposed for B-line identification is challenging for three reasons. First, the investigated tasks are related but different (e.g., B-line detection, COVID-19 severity scoring, differential diagnosis). Second, data preprocessing decisions vary across studies. Third, datasets and model implementations are not available publicly.

To this end, we present a benchmark of deep learning approaches for B-line detection in LUS videos. We formulate B-line detection as a classification task at three levels -- at the level of multi-frame clips, at the level of single frames, and at the level of individual pixels (i.e., segmentation) -- to study the benefit of temporal information and annotations. Predictions are aggregated to detect B-lines in full LUS videos. Specifically for the pixel-level, we propose a novel approach to B-line localization, predicting the landmark locations as single points where B-lines originate from the pleura. This approach addresses the need for interpretable B-line detection and quantification at a relatively low annotation cost. To facilitate further research into computer-assisted LUS interpretation, we make publicly available our expert-annotated dataset.

\section{Related Work}
\label{sec:related}
In this section, we discuss prior work on automated detection and localization of B-lines, as well as closely related methods for severity assessment of COVID-19 and differential diagnosis in LUS.
 
Brattain\etal \cite{brattain2013automated} transformed curvilinear LUS frames from polar coordinates to a Cartesian grid, in which B-lines appear as vertical lines. The detection was performed column-wise using handcrafted intensity-based features. Moshavegh\etal \cite{moshavegh2018automatic} used classical image analysis techniques for segmentation of B-lines. Brusasco\etal \cite{brusasco2019quantitative} developed an automatic scoring method based on the percentage of the pleura from where B-lines originate, which showed a strong association with the measured extravascular lung water. Alternatively, Anantrasirichai\etal \cite{anantrasirichai2017line} proposed a procedure for localizing B-lines in LUS frames from kidney failure patients by solving line detection as an inverse problem using the Radon transform. Karaku\c{s}\etal \cite{karakucs2020detection} extended this method for B-line localization in LUS frames from COVID-19 patients. The comparatively long computation time of this method may, however, be a limitation for clinical translation.

More recent work has mainly focused on the application of deep learning approaches. Van Sloun and Demi~\cite{van2019localizing} applied a convolutional neural network (CNN) for B-line detection in single LUS frames, using gradient-based class activation mapping (Grad-CAM)~\cite{selvaraju2017grad} as a form of weakly supervised localization. The network was trained on frames from a collection of both phantom and patient LUS videos. 
Building upon this work, Roy\etal~\cite{roy2020deep} used a spatial transformer network (STN) to extract salient crops from the input frame, serving both as a weak localization method and as input to a soft ordinal regression network for COVID-19 severity scoring. Predicted severity scores for single frames were aggregated using a uninorm function to obtain score predictions per video. Segmentation of COVID-19 imaging biomarkers was additionally investigated, but not pursued for score prediction. The models were developed and evaluated based on 277 LUS videos from 35 patients in the ICLUS dataset.
Frank\etal~\cite{frank2021integrating} achieved improved performance on the same dataset by incorporating anatomical features and LUS artifacts as additional input channels to a CNN, which were extracted using classical image analysis techniques.
Mason\etal~\cite{mason2021lung} investigated B-line segmentation in LUS frames of COVID-19 and community-acquired pneumonia patients. The authors reported improved segmentation performance when leveraging domain adaptation. To address the challenge of acquiring expert annotations, Zhao\etal~\cite{zhao2022covid} investigated the use of simulated lung ultrasound images for training CNNs to segment B-line artifacts. 
Tan\etal~\cite{tan2022automated} used an instance segmentation approach to localize B-lines in single frames. Quantification of distinct B-lines for entire LUS videos was achieved using a tracking algorithm. The automated B-line counts were shown to correlate well with expert counts, and to correlate moderately with the fluid overload for dialysis patients.
Alternatively, Kulhare\etal~\cite{kulhare2018ultrasound} proposed a single-shot CNN for detection and localization of LUS characteristics, including B-lines, in LUS frames. However, since all data were acquired from animals with induced pathology, model generalization to human LUS videos remains to be investigated. 

Others have presented methods that leverage both spatial and temporal information. Kerdegari\etal \cite{kerdegari2021b} used a CNN for feature extraction from single frames followed by a long short-term memory (LSTM) unit for B-line detection in LUS videos. Both spatial and temporal attention mechanisms were used, which provide weakly supervised spatial and temporal localization of B-lines in the video. The experiments were performed on a private dataset of close to 3600 videos from 60 severe dengue patients. 
Baloescu\etal~\cite{baloescu2020automated} trained 2D and 3D CNNs on short LUS clips for B-line detection as well as severity assessment. The models were developed and evaluated on a private dataset of 400 videos, each from a unique patient. Whereas severity score prediction was shown to be more challenging, the authors reported strong performance on the binary B-line detection task using frames of 75$\times$75 pixels as input size.
Born\etal~\cite{born2021accelerating} compared both frame- and clip-level CNNs for classification of COVID-19, bacterial pneumonia, and healthy LUS images. Predictions by the networks were visually examined using class activation mapping (CAM)~\cite{zhou2016learning}. The authors collected a heterogeneous set of 202 LUS videos and 59 still frames from 216 patients for the experiments, which was made available to the public as the POCOVID dataset. On a prior version of this dataset, Awasthi\etal\cite{awasthi2021mini} benchmarked several lightweight CNNs for embedded applications.

\section{Contributions}
To summarize, the major contributions of this work are:
\begin{itemize}
    \item We present a benchmark of deep neural network architectures for B-line detection in LUS videos. The presence of B-lines is predicted at the level of multi-frame clips, single frames, and individual pixels (i.e. segmentation), to investigate the benefit of temporal information and annotations. The area under the receiver operating characteristic curve ranges from 0.864 to 0.955 for all evaluated methods. The best performance is achieved by clip-level models. 

    \item We propose a novel formulation of B-line localization through landmark detection. In this approach, segmentation networks are trained using only single-point annotations located where B-lines originate from the pleura, which has a relatively low annotation cost. The best networks achieve an F$_{1}$-score of 0.65, performing comparably to the inter-observer agreement.

    \item We publicly release a new LUS dataset comprising 1,419 videos from 113 patients, along with expert annotations. All videos are labeled as positive or negative for the presence of B-lines. For the positive videos, a total of 15,755 B-lines are annotated on 10,371 frames.
\end{itemize}

\section{Dataset}
We refer to the data collected and curated in this study as the \textit{Boston Emergency Department Lung Ultrasound (BEDLUS)} dataset. All data were collected from patients admitted to the emergency department of Brigham and Women's Hospital between November 2020 and March 2021 with symptoms suggestive of a flu-like illness including shortness of breath, cough, fever, sore throat, myalgias, or fatigue. The study was performed in accordance with protocols approved by the IRB of Brigham and Women's Hospital (2020P002367, approved on August 31, 2020; 2021P001446, approved on February 16, 2022). 
A total of 113 patients were examined, yielding 1,419 LUS videos comprising 188,670 frames. The mean patient age was 60.6 years and 55\% were female. The discharge diagnoses of 55\% of the patients are traditionally known to cause B-lines, which include heart failure exacerbation, interstitial lung disease, and bacterial or viral pneumonia (e.g., due to COVID-19 infection). The remaining 45\% of the patients had discharge diagnoses that are not conventionally associated with B-lines.

Patient LUS examinations were performed according to the BLUE protocol~\cite{lichtenstein2008relevance}. In brief, LUS videos were acquired at two anterior, two lateral, and two posterior zones of the patient's chest for both the left and right lung, typically resulting in 12 videos per patient examination. In each zone, the transducer was positioned roughly perpendicular to the ribs at an intercostal space. Videos were recorded while the transducer was held stationary for six seconds to contain at least one respiratory cycle. For some patients not all zones were imaged due to patient discomfort, positioning, or instability, whereas for other patients multiple LUS videos of the same zone were collected. All videos were acquired at frame rates between 15-46~Hz using a low-frequency (1-5~MHz), curvilinear transducer (Mindray North America, New Jersey, USA) without predetermined settings for the depth, focal point, and gain. No image-enhancement modes were used for acquisition. 

The videos were downloaded in DICOM format from the Picture Archiving and Communication System (PACS) (Visage 7, Promedicus Limited, San Diego, USA) at Brigham and Women's Hospital. The DICOM files were first converted to MP4 format using an open-source medical image viewer\footnote{Horos, https://horosproject.org} and subsequently de-identified using a free software package\footnote{Clip Deidentifier, https://www.ultrasoundoftheweek.com/clipdeidentifier}. 
After de-identification, all LUS video frames were preprocessed by cropping the ultrasound window to remove embedded information, padding the crop to the original aspect ratio, downsampling the image to a size of 384$\times$256 pixels (width by height), and normalizing the intensity values to the range of $[0,1]$. 

\begin{figure}[t]
    \centering
    \includegraphics[width=0.98\columnwidth]{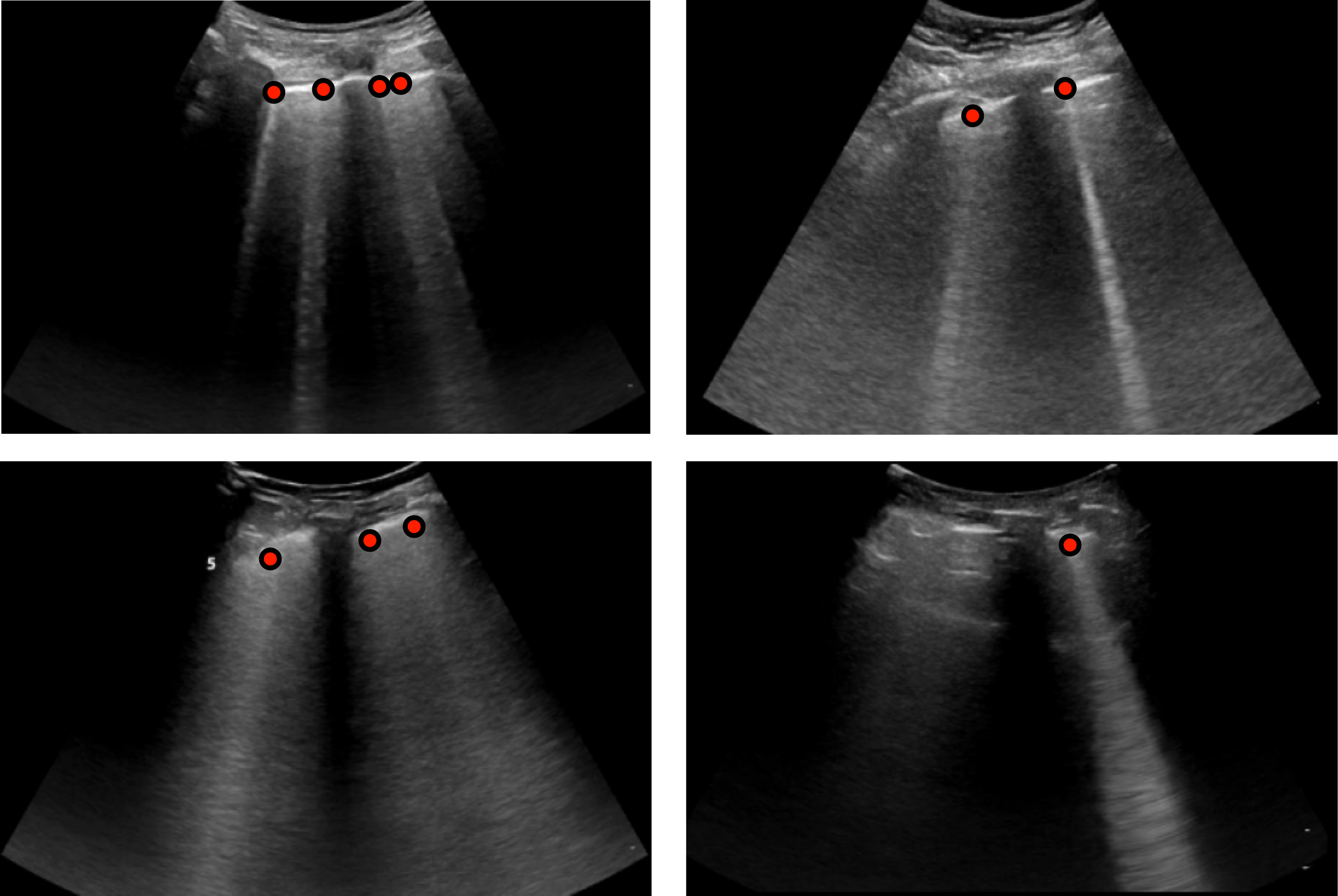}
    \caption{Example LUS frames from the curated dataset with corresponding expert annotations of B-line origins as red points.}
    \label{fig:frames}
\end{figure}

The LUS videos were first labeled based on a consensus decision by two lung ultrasound experts (N.D. and A.G.) indicating the presence of at least one B-line throughout the video. All 719 videos (50.7\%) that were positively labeled for the presence of B-lines were subsequently presented to one of the experts for B-line annotation. The expert (N.D.) annotated all B-lines in the presented frames by locating the points where each B-line originates from the pleura. Four example frames with B-lines including the corresponding point of origin annotations are shown in Fig.~\ref{fig:frames}. To account for annotator workload and because successive frames have a strong similarity, only every fourth frame of the positively labeled videos was presented for annotation. This resulted in a total of 15,755 annotated B-line origins on 10,371 frames. Labeling and annotating was performed using the DiagnosUs software application\footnote{DiagnosUs, https://www.diagnosus.com}. The dataset and trained model parameters are available via \UrlFont\url{https://www.dropbox.com/s/bvha488i2eq2fek/BEDLUS\%20data\%20instructions.pdf?dl=0}.

\begin{figure*}[th!]
    \centering
    \includegraphics[width=0.98\textwidth]{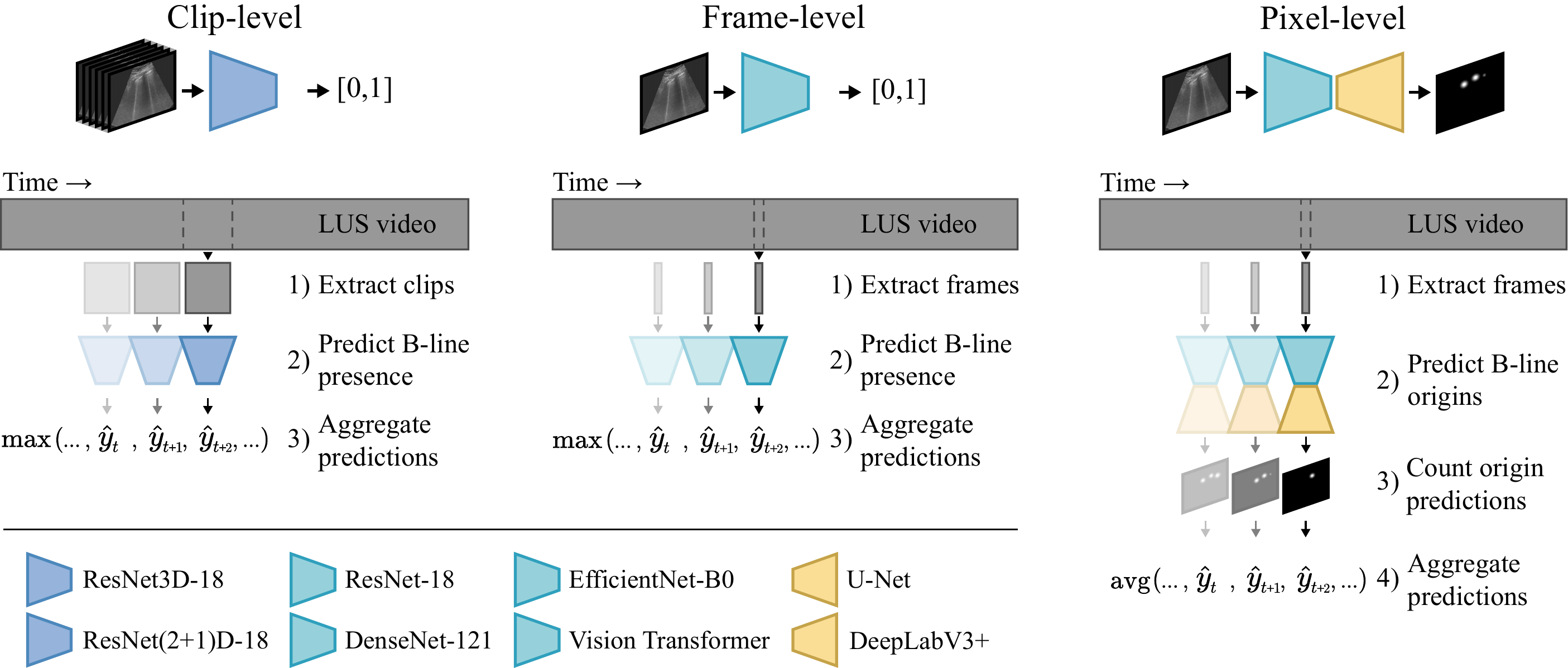}
    \caption{Overview of the benchmarked methods. We investigate B-line detection in LUS videos by aggregating classification predictions at the level of multi-frame clips, single frames, and individual pixels. At the pixel-level, isolated detections are counted by thresholding the predicted probability heatmap as an intermediate step before aggregation.}
    \label{fig:overview}
\end{figure*}

\section{Method}
\subsection{Problem Formulation}
For the task of B-line detection, we are provided with a LUS video composed of a sequence of frames $X = (x_t)_{t=1}^{T}$, where $x_t \in \mathcal{X} \subset \mathbb{R}^{H \times W}$ is a single frame of height $H$ and width $W$ on position $t$ in time. The videos vary in duration and thus $T$ is not fixed. 
We denote $\mathcal{V}$ as the input space of LUS videos, $\mathcal{X}$ as the input space of LUS frames, and $\mathcal{Y} = [0,1]$ as the range of the predicted probability for the presence of B-lines. In this work, we compare different mappings $f:\mathcal{V} \rightarrow \mathcal{Y}$, predicting B-line presence for entire LUS videos.

We investigate three types of mappings for B-line detection -- clip-level, frame-level, and pixel-level -- by training frequently used and state-of-the-art deep neural network architectures on single frames or short clips extracted from the videos. The corresponding predictions are aggregated to obtain a prediction for the full LUS video. An overview of the three approaches is illustrated in Fig.~\ref{fig:overview}. As for aggregation, we compared: 1) averaging all predictions, 2) selecting the maximum prediction, or 3) selecting the maximum of a moving average prediction. For each of the three levels, the aggregation method that achieved the best performance on the validation set was selected. 

\subsubsection{Clip-level} To predict the presence of B-lines for a multi-frame clip, the mapping $f_{clip}: \mathcal{V}^L \rightarrow \mathcal{Y}$ is learned, where $\mathcal{V}^L \subset \mathcal{V}$ is the subset of clips from LUS videos with $L$ successive frames. To model $f_{clip}$, a ResNet3D-18 and a ResNet(2+1)D-18~\cite{tran2018closer} were trained starting from network parameters pretrained on the Kinetics-400 dataset. To compare to related work, a 3D U-Net encoder with a classification head was trained starting from network parameters pretrained on a medical dataset\cite{zhou2021models}, similar to Born\etal\cite{born2021accelerating}. A clip length of $L=16$ was selected as a trade-off between incorporated temporal information and computational cost. The clip-level results were aggregated to obtain a prediction for the entire video by selecting the maximum predicted probability for the successive clips.  

\subsubsection{Frame-level} To predict the B-line presence for single frames, the mapping $f_{frame}:\mathcal{X} \rightarrow \mathcal{Y}$ is learned. We evaluated ResNet-18~\cite{he2016deep}, DenseNet-121~\cite{huang2017densely}, EfficientNet-B0~\cite{tan2019efficientnet}, and the Vision Transformer~\cite{dosovitskiy2020vit} (variant: tiny; patch size: 16$\times$16 pixels)~\cite{touvron2021training}, all initialized with ImageNet pretrained parameters, for learning the mapping $f_{frame}$. No improvements to the performance on the validation set were observed for larger network variants. For comparison to the state of the art in automated LUS assessment, we additionally evaluated VGG16 with a custom classification head from Born\etal\cite{born2021accelerating}, the STN+CNN from Roy\etal\cite{roy2020deep}, and MobileNetV2 as benchmarked in Awasthi\etal\cite{awasthi2021mini}, all trained on our dataset. Predictions for entire videos were obtained by selecting the maximum predicted probability for the individual frames.

\subsubsection{Pixel-level} To predict the presence of B-lines at each pixel in a frame, the mapping $f_{pixel}:\mathcal{X} \rightarrow \mathcal{Y}^{H \times W}$ is learned. More specifically, we propose to localize only the locations where B-lines originate from the pleura. This task can therefore be considered as landmark detection using a segmentation network. To model $f_{pixel}$, we used ResNet-18, DenseNet-121, and EfficientNet-B0 as pretrained encoders, coupled with the decoder structure of U-Net~\cite{unet} (five upsampling steps with 256 filters after the first upsample operation) and DeepLabV3+~\cite{chen2018encoder}. Decoder parameters were randomly initialized using He normal initialization~\cite{he2015delving}. For comparison to related work, we also trained and evaluated a U-Net similar to Mason\etal\cite{mason2021lung}. Different from the previous two levels, the LUS video prediction was calculated as the average number of detections per frame. To quantify the detected locations in each frame, the predicted probability heatmap was first binarized using a threshold of 0.5, followed by counting the number of isolated components.

\subsection{Network Training}
\label{sec:net_train}
The dataset was randomly split on a patient-level into a set for model development and a set for testing. The development set contained 80\% of the patients (1,112 videos with B-lines present in 49.4\%), which was further subdivided into five folds of 18 unique patients for cross-validation. Training was repeated five times for all architectures, each using a different fold for validation and the remaining four folds for training. The other 20\% of the patients (307 videos with B-lines present in 55.4\%) were allocated as a test set for a final independent evaluation of the trained models. It should be noted that out of the full dataset of 1,419 LUS videos, 719 contained B-lines based on expert assessment (A.G. and N.D.) and 700 did not.

Batches of data were randomly sampled during training. Each batch was half-filled with annotated frames. The remaining half was sampled from all negatively labeled LUS videos. One epoch was defined as a pass over all annotated frames. Since the number of extracted frames from each LUS video differs, and because of the similar appearance of successive frames, training could be biased towards the videos with more available frames. 
Hence, we correct for the possible bias by weighting each frame in the training set by $\omega$, the reciprocal of the total number of frames used from that particular LUS video (e.g., in case five frames are used from one LUS video, each frame gets a weighting of $\omega = \frac{1}{5}$ during training). All weights of the negative frames were multiplied by a correction factor to balance the contribution of positive and negative frames in each epoch. The same procedure was followed for training the clip-level networks. However, instead of sampling individual frames, batches were half-filled with 16-frame clips of which at least one frame was annotated. The other half was sampled from all possible 16-frame clips that are part of negatively labeled LUS videos in the training set.

We built upon pretrained networks that were implemented in the Pytorch framework~\cite{Pytorch}. Networks were trained for a maximum of 100 epochs using the Adam optimization algorithm ($\beta_{1}=0.9$, $\beta_{2}=0.999$). If the validation loss did not improve for five consecutive epochs, the learning rate was halved. We also employed an early stopping policy, which stops training after ten consecutive epochs without improvements to the validation loss. Network parameters from the last epoch with validation loss improvements were saved. 
Data augmentation was performed on the fly by randomly applying translations, rotations, scaling, left-right flips, and occlusions. Additionally, we applied random brightness and contrast changes, Gaussian noise, and Gaussian blurring to improve the robustness against the acquisition settings.
Furthermore, since some networks were pretrained on natural images and therefore require a 3-channel input, the same grayscale ultrasound image was duplicated for each color channel. Hyperparameters were tuned based on preliminary experiments on the validation set. Level-specific details are further elaborated below. 

\subsubsection{Clip-level}
Classification networks for clips were trained using a learning rate of $1\times10^{-4}$ by minimizing the binary cross-entropy loss defined as:
\begin{equation}
\mathcal{L}_{BCE} = -\omega\,(y\ln(\hat{y}) + (1-y)\ln(1-\hat{y})),
\label{Eq:BCE}
\end{equation}
where $y$ is the class label, $\hat{y}=f_{clip}(X)$ is the predicted probability for the positive class (i.e., the presence of B-lines), and $\omega$ is a normalizing weight to balance the effect of clips extracted from videos with variable length.
Due to the large clip size and memory constraints, we used four clips per batch during training.

\subsubsection{Frame-level}
Networks for classification of single frames were also trained by minimizing the binary cross-entropy loss in Eq.~\ref{Eq:BCE}, except here $\hat{y}=f_{frame}(x)$ is the predicted probability of B-line presence for a single frame $x$. The learning rate was $2\times10^{-6}$ for training ResNet-18, DenseNet-121, MobileNetV2, and the Vision Transformer, $1\times10^{-5}$ for training EfficientNet-B0 and VGG16, and $5\times10^{-5}$ for training the STN+CNN. A batch size of 32 was used to train all networks.
For the Vision Transformer, images were zero-padded to the size of 384$\times$384 pixels as required for the pretrained implementation. Moreover, preliminary results on the validation set indicated the benefit of stronger data augmentation for the Vision Transformer in contrast to the other networks. This is in line with prior research \cite{touvron2021training} and can be attributed to the lower inductive bias of the architecture.

\subsubsection{Pixel-level}
Segmentation networks for pixel-level prediction were trained using a learning rate of $1\times10^{-3}$ by minimizing a combination of the focal loss and Dice loss. The focal loss~\cite{lin2017focal} is an extension of the binary cross-entropy, parameterized by $\gamma$. Increasing $\gamma$ reduces the relative importance of well-classified pixels, which can alleviate class imbalance by shifting the focus away from the large number of correctly classified background pixels.
The focal loss was defined as:
\begin{equation}
\begin{split}
\mathcal{L}_{F} = -\omega\,(\beta(1-\hat{y})^{\gamma}y\ln(\hat{y})
+ \hat{y}^{\gamma}(1-y)\ln(1-\hat{y})),
\end{split}
\end{equation}
where $y$ is the class label map, $\hat{y}=f_{pixel}(x)$ is the predicted probability map, and $\beta$ is a foreground weight to further combat the strong class imbalance. 
The Dice loss was defined as:
\begin{equation}
\mathcal{L}_{DSC} = \omega\left(1-\frac{2\,y\,\hat{y}+\epsilon}{y+\hat{y}+\epsilon}\right),
\end{equation}
where $y$ is the class label map, $\hat{y}=f_{pixel}(x)$ is the predicted probability map, and $\epsilon = 1$ is a smoothing term to make the loss differentiable and to avoid zero in the numerator.
The combined segmentation loss was defined as:
\begin{equation}
\mathcal{L}_{SEG} = \alpha\,\mathcal{L}_{DSC} + (1-\alpha)\,\mathcal{L}_{F},
\end{equation}
where $\alpha$ determines the contribution of each loss component. Based on a hyperparameter grid search, we found that a configuration of $\alpha = \frac{2}{3}$, $\gamma = 2$, and $\beta = 50$ resulted in the best performance on the validation set. Similar to the frame-level training, a batch size of 32 frames was used.

To generate label maps for training the segmentation networks, coordinates of the annotated B-line origins were converted into circles with a 4~mm diameter (i.e., between 5 and 17 pixels depending on the spatial size of a pixel for a particular video). This size was selected based on the observations that smaller diameters resulted in unstable training, whereas larger diameters led to circles that more frequently overlapped. All code for data preprocessing and the experiments is available at \url{https://github.com/RTLucassen/B-line_detection}.

\begin{table*}[t]
\centering
\caption{Benchmark of architectures at clip-level, frame-level, and pixel-level for predicting B-line presence on test set LUS videos. Best AUC results at each level are in bold.}
\label{tab:level_benchmark}
\begin{tabular}{lllrrcccc}
\toprule
Level       & \multicolumn{3}{c}{Architecture} & Related Work\hspace{0.2cm} & \multicolumn{2}{c}{Mean $\pm$ SD}                    & \multicolumn{2}{c}{Ensemble}      \\ 
\cmidrule(lr){2-4}\cmidrule(lr){6-7}\cmidrule(lr){8-9}  
            & Encoder               & Decoder        & \hspace{-0.3cm}Parameters &  & F$_{1}$-score     & AUC                             & F$_{1}$-score & AUC   \\
\midrule
Pixel       & U-Net                  & U-Net         & 7.9M  & Mason\etal\cite{mason2021lung}         & 0.857 $\pm$ 0.006 &          0.913 $\pm$ 0.009     & 0.865 &          0.922 \\
            & ResNet-18              & U-Net         & 14.3M & --                                     & 0.843 $\pm$ 0.023 &          0.899 $\pm$ 0.013     & 0.861 &          0.919 \\
            & ResNet-18              & DeepLabV3+    & 12.3M & --                                     & 0.869 $\pm$ 0.009 &          0.902 $\pm$ 0.011     & 0.886 &          0.911 \\
            & DenseNet-121           & U-Net         & 13.6M & --                                     & 0.861 $\pm$ 0.015 &          0.911 $\pm$ 0.007     & 0.880 &          0.920 \\
            & EfficientNet-B0        & U-Net         & 6.3M  & --                                     & 0.875 $\pm$ 0.020 & \textbf{0.918} $\pm$ 0.015     & 0.890 & \textbf{0.930} \\
            & EfficientNet-B0        & DeepLabV3+    & 4.9M  & --                                     & 0.877 $\pm$ 0.007 &          0.912 $\pm$ 0.010     & 0.886 &          0.922 \\ \\
                      
Frame       & \multicolumn{2}{l}{MobileNetV2}        & 2.2M  &  Awasthi\etal\cite{awasthi2021mini}    & 0.847 $\pm$ 0.007 &          0.911 $\pm$ 0.003     & 0.837 &          0.917 \\
            & \multicolumn{2}{l}{STN+CNN}            & 1.9M  &  Roy\etal\cite{roy2020deep}            & 0.837 $\pm$ 0.028 &          0.907 $\pm$ 0.018     & 0.848 &          0.917 \\
            & \multicolumn{2}{l}{VGG16}              & 14.9M &  Born\etal\cite{born2021accelerating}  & 0.847 $\pm$ 0.019 &          0.907 $\pm$ 0.016     & 0.860 &          0.923 \\
            & \multicolumn{2}{l}{ResNet-18}          & 11.2M & --                                     & 0.848 $\pm$ 0.012 &          0.914 $\pm$ 0.007     & 0.867 &          0.919 \\
            & \multicolumn{2}{l}{DenseNet-121}       & 7.0M  & --                                     & 0.851 $\pm$ 0.015 &          0.914 $\pm$ 0.014     & 0.862 &          0.923 \\
            & \multicolumn{2}{l}{EfficientNet-B0}    & 4.0M  & --                                     & 0.850 $\pm$ 0.017 & \textbf{0.922} $\pm$ 0.007     & 0.876 &          0.929 \\
            & \multicolumn{2}{l}{Vision Transformer} & 5.6M  & --                                     & 0.849 $\pm$ 0.012 &          0.918 $\pm$ 0.005     & 0.868 & \textbf{0.930} \\ \\
                     
Clip        & \multicolumn{2}{l}{3D U-Net}           & 7.6M  & Born\etal\cite{born2021accelerating}   & 0.782 $\pm$ 0.006 &          0.840 $\pm$ 0.005     & 0.805 &          0.864 \\
            & \multicolumn{2}{l}{ResNet3D-18}        & 33.2M & --                                     & 0.884 $\pm$ 0.017 & \textbf{0.947} $\pm$ 0.011     & 0.896 & \textbf{0.955} \\
            & \multicolumn{2}{l}{ResNet(2+1)D-18}    & 31.3M & --                                     & 0.878 $\pm$ 0.009 &          0.938 $\pm$ 0.002     & 0.905 &          0.951 \\
\bottomrule
\end{tabular}
\end{table*}

\begin{figure*}[t]
    \begin{flushright}
    \includegraphics[width=0.97\textwidth]{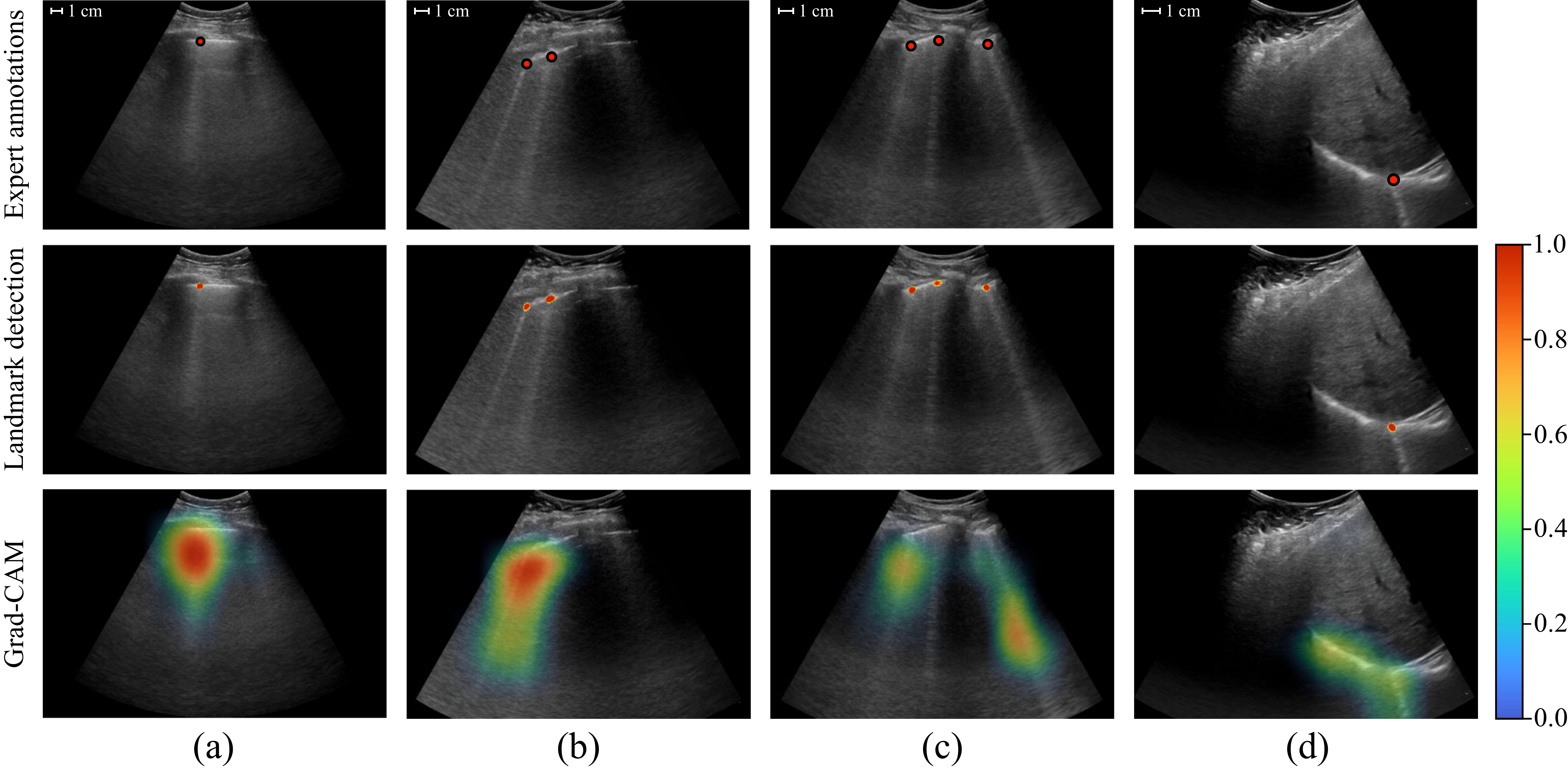}
    \caption{Comparison between single point predictions from EfficientNet-B0 with U-Net (middle row) and class activation maps by Grad-CAM based on EfficientNet-B0 (bottom row) for selected example frames from test set videos. The corresponding expert annotations are shown in the upper row. Segmentation networks for landmark detection are trained to detect the originating locations of B-lines on the pleura, whereas Grad-CAM highlights the maximally activating regions for the B-lines class. The landmark detection heatmap was masked using a threshold of 0.5, similar to how the detections were counted. Class activation maps were normalized based on the validation set. The predicted B-line origins provide indications of the located B-lines that are more specific and less ambiguous in comparison to the indications provided by Grad-CAM.}
    \label{fig:localization}
    \end{flushright}
\end{figure*}

\subsection{Inference}
After training, the networks were used to predict the presence of B-lines for all LUS videos in the test set. For the clip-level networks, each video was subdivided into sequential 16-frame clips. The predicted probabilities for all clips were then aggregated by selecting the maximum predicted probability as the prediction for the video. For the frame-level networks, the predicted probabilities for all single frames were similarly aggregated by selecting the maximum predicted probability to obtain a video prediction. In contrast, the pixel-level segmentation networks output a heatmap with high probabilities near predicted B-line origins and low predicted probabilities elsewhere. To localize the predicted origins, the heatmap was binarized using a threshold of 0.5 and the centroid was computed for all isolated points. To arrive at a B-line detection score for a LUS video, the number of predicted B-line origins was averaged over all video frames.

Moreover, ensemble predictions were calculated by first averaging the predictions of the five network instances trained with cross-validation for all single frames or multi-frame clips, followed by aggregating the averaged predictions to obtain the ensemble prediction for a LUS video.

Finally, we performed an exploratory analysis of approaches to leverage model predictions across the three levels. Using the binary detection decisions after the classification thresholds have been applied to the model predictions, we investigated majority voting and reporting only unanimous detection decisions.

\section{Results}
\subsection{Detection of B-Lines}
The B-line detection performance on the previously unseen test set was expressed as the F$_{1}$-score and the area under the receiver operating characteristic (ROC) curve, abbreviated as AUC. The F$_{1}$-score is based on the classification threshold that balanced the precision and recall on the validation set. For each architecture, five network instances were trained using cross validation. In Table~\ref{tab:level_benchmark}, we report the mean and standard deviation (SD) of the F$_{1}$-score and AUC for the five network instances. In addition, predictions by the five networks were averaged before aggregation to obtain ensemble results for each architecture. 
 
As shown in Table~\ref{tab:level_benchmark}, the ensemble AUC ranges from 0.864 to 0.955 for all evaluated methods. Within this range, the best B-line detection performance was achieved by ResNet3D-18 at the clip-level with a mean AUC of 0.947, and an ensemble AUC of 0.955. Both clip-level ResNets outperformed all frame- and pixel-level networks. EfficientNet-B0 with the U-Net decoder performed on par with the best frame-level models as measured by the AUC, but reached a better F$_{1}$-score. Differences in terms of performance between the architectures at each level are relatively small, except for the 3D U-Net encoder with classification head, which performed considerably worse than the clip-level ResNets. For all model architectures, the ensemble of five network instances performed better than the individual networks on average. The combination of DenseNet-121 and DeepLabV3+ is not featured in Table~\ref{tab:level_benchmark}, as atrous convolutions for pooling are not directly applicable to the DenseNet-121 encoder.

Moreover, two approaches for combining models across the three levels were explored. To maintain a fair evaluation, we selected the best model for each level based on the validation results, which were EfficientNet-B0 with U-Net, ResNet-18, and ResNet3D-18. By combining the detection decisions using majority voting, an F$_{1}$-score of 0.898 on the test set was obtained. Alternatively, the three models reached a unanimous detection decision for 81.1\% of the videos in the test set. On this subset, an F$_{1}$-score of 0.955 was achieved. Note that for the remaining 18.9\% of the videos, no prediction was reported.

\subsection{Single-Point Localization of B-Lines }
The B-line localization performance of the pixel-level networks was quantitatively evaluated based on the distance between the annotated and predicted B-line origins on every fourth frame of all test set videos with B-line annotations. A detected location was considered to be a true positive if the Euclidean distance to any of the annotated origins was less than 5~mm. This threshold distance was deemed suitable, as the distance between 99\% of all possible pairs of annotated B-line origins in the dataset exceeds 5~mm. 
In case that more than one detection was found within the detection distance from an annotated origin, only one contributed to the true positive count. All annotated B-line origins without detections present within a distance of 5~mm were considered to be false negatives and all detections that were not within 5~mm of any annotated origin location were counted as false positives. The precision and recall were calculated based on the sum of the true positive, false positive, and false negative detections for each LUS video. The localization performance was expressed as the F$_{1}$-score, calculated as the harmonic mean of the average precision and recall over all videos with annotations in the test set.

The localization results in Table~\ref{tab:localization} show the performance of the pixel-level network ensembles in terms of the precision, recall, and F$_{1}$-score. EfficientNet-B0 with the U-Net decoder achieved the best performance with an F$_{1}$-score of 0.65 on the annotated test set videos.
The inter-observer agreement was determined on a randomly selected subset of two positive test set videos per patient. The selected videos were annotated by a second expert (A.G.) using an identical annotation set up. The $F_{1}$-score between the annotations of the two experts was 0.63. On the selected subset of videos, EfficientNet-B0 with the U-Net decoder achieved an $F_{1}$-score of 0.63 when evaluated against the initial expert (N.G.) that also provided the training annotations, and achieved an $F_{1}$-score of 0.57 when evaluated against the second expert (A.G.). 

\begin{table}[b]
\centering
\caption{Single-point localization performance for pixel-level network ensembles on the annotated test set LUS videos\\ using a detection distance of 5~mm.}
\label{tab:localization}
\begin{tabular}{@{}llcccc@{}}
\toprule
\multicolumn{2}{c}{Architecture} & Precision & Recall & F$_{1}$-score\\ 
\cmidrule(lr){1-2}
Encoder           & Decoder & & & \\ \midrule
U-Net             & U-Net          & 0.57          & 0.66          & 0.61\\
ResNet-18         & U-Net          & 0.59          & 0.62          & 0.61\\
ResNet-18         & DeepLabV3+     & 0.60          & 0.65          & 0.63\\
DenseNet-121      & U-Net          & 0.57          & 0.64          & 0.60\\
EfficientNet-B0   & U-Net          & 0.61          & 0.70          & 0.65\\ 
EfficientNet-B0   & DeepLabV3+     & 0.62          & 0.67          & 0.64\\ \bottomrule
\end{tabular}
\end{table}

The localization performance of the single-point detection approach was qualitatively compared to that of class activation maps, constructed using a classification network with Grad-CAM \cite{selvaraju2017grad}. The class activation maps highlight image regions that contributed most to the predicted probability for a particular class. A visual comparison between landmark detection predictions (from EfficientNet-B0 with U-Net) and class activation maps by Grad-CAM (based on EfficientNet-B0) for selected example frames from the test set is shown in Fig.~\ref{fig:localization}. 
Although the class activation maps highlight the regions containing B-lines, because of the lower resolution, the method appears to be less suitable for differentiating individual B-lines (e.g., Fig.~\ref{fig:localization}b~\&~c) and occasionally results in ambiguity (e.g., Fig.~\ref{fig:localization}d). The same limitations were observed for Grad-CAM in combination with clip-level networks.

\section{Discussion and Conclusion}
In this work, we benchmarked deep learning approaches for classification of B-line presence in LUS videos at the level of multi-frame clips, single frames, and individual pixels (i.e., segmentation). For the pixel-level, we proposed a novel formulation of B-line localization, predicting the origin locations as single points.

The benchmark results in Table~\ref{tab:level_benchmark} show that clip-level models reached the best B-line detection performance in terms of the AUC and F$_{1}$-score. This suggests the benefit of temporal information for detecting patterns of B-line artifacts in LUS videos. Differences between network architectures at each level are relatively small, with the exception of the 3D U-Net encoder for clip-level predictions. This in line with the observations from Born\etal\cite{born2021accelerating}, and can likely be explained by the comparatively shallow depth of the architecture for classification. Besides the detection performance, interpretability is an important aspect of machine learning models for safety-critical domains such as medical image analysis \cite{rudin2019stop}. Although the best frame- and pixel-level models performed comparably, an advantage of the pixel-level models is that the predicted B-line origin locations directly contribute to the video classification score in a human-understandable way, making the classification prediction more easily interpretable than for the frame- and clip-level models. Inference time is another factor to consider for clinical translation. Using a system with an Intel Core i7-6800K CPU, 16GB of RAM, and an Nvidia TITAN Xp GPU, the mean inference time for a video of average length in our dataset (133 frames) was measured at less than one second for all frame-level models, and between 1.0 to 1.5 seconds for all pixel- and clip-level models. Since ensembled networks are computationally independent and can therefore run in parallel, given sufficient computational resources, similar runtimes could be achieved for the ensembles.

Combining binary detection decisions from model ensembles across the three levels using majority voting did not lead to an improvement in the F$_{1}$-score, while increasing the computational cost. Alternatively, by considering agreement among the three models as a requirement, a unanimous detection decision was reported for 81.1\% of the test set videos. The F$_{1}$-score on this subset was substantially higher than for the separate model ensembles on the complete test set. It should be noted that the F$_{1}$-score is approximately the same for any large enough subset of the test set selected at random. Hence, the increase in the F$_{1}$-score suggests that this approach can automatically identify and filter out error-prone LUS videos better than random chance. Because this approach does not report a detection decision for all videos, the utility may differ depending on the application.

The best single-point localization performance with an F$_{1}$-score of 0.65 was achieved by the ensemble of EfficientNet-B0 with the U-Net decoder. The agreement between the initial expert and the model predictions as measured by the F$_{1}$-score is comparable to the agreement between the initial and second expert. The second expert agreed with the model in slightly fewer cases, which can likely be explained by the fact that the model was not trained using annotations from this expert. Van Sloun and Demi \cite{van2019localizing} used class activation maps produced by Grad-CAM as a method for weakly supervised localization. A visual comparison between class activation maps and B-line origin predictions is shown in Fig.~\ref{fig:localization}. The predicted origins provide indications of located B-lines that are more specific and less ambiguous in comparison to the class activation maps, making the single-point approach more suitable as a method for localizing individual B-lines in images. As for network training, even though acquiring single-click annotations of B-line origins takes more effort than frame or clip labeling, the process is less laborious than delineating entire B-line regions in images \cite{roy2020deep, mason2021lung}.

\begin{figure}[t]
\centering
\includegraphics[width=0.97\columnwidth]{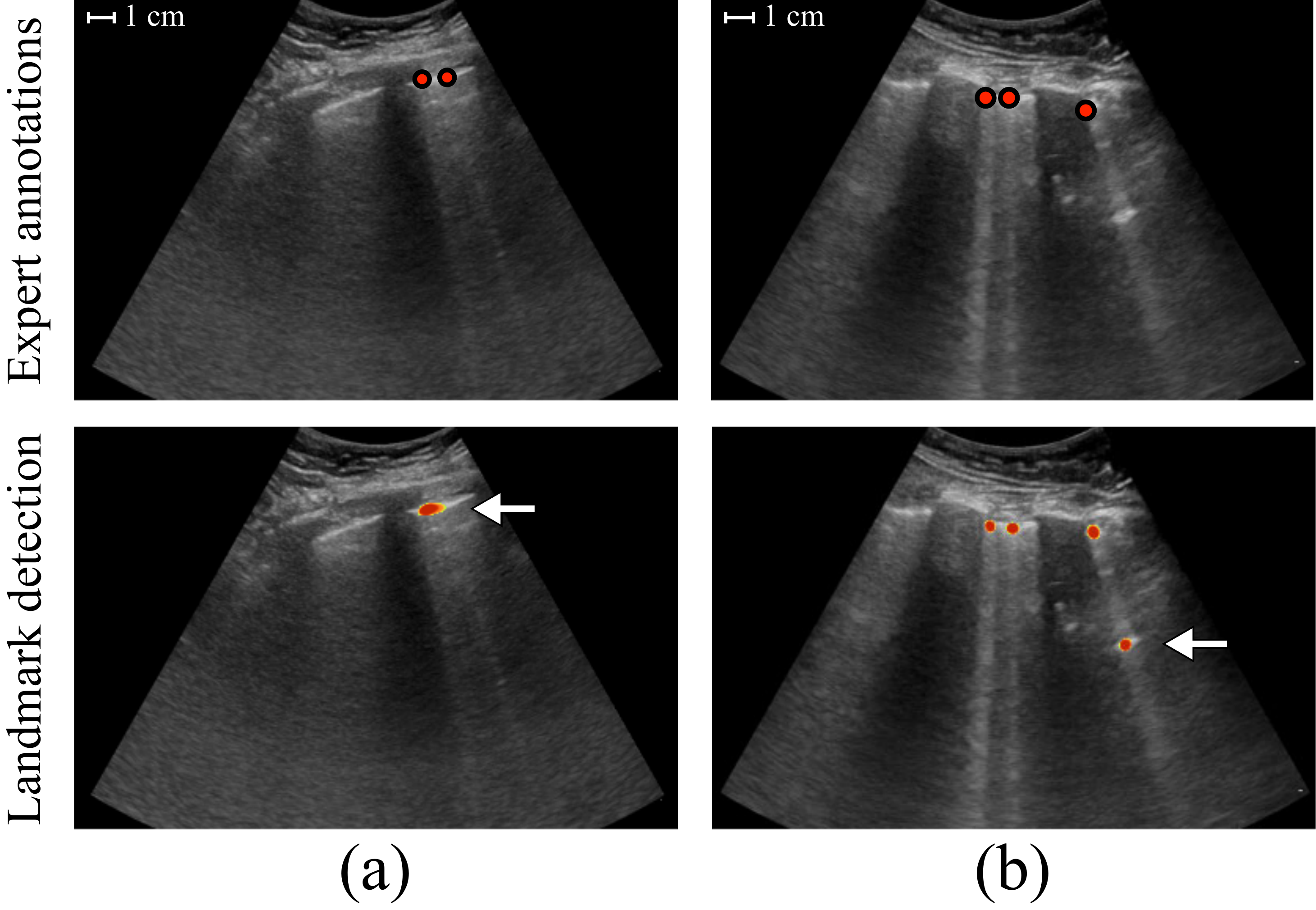}
\caption{Two example frames with incorrectly predicted origins indicated by the arrows. (a) The predictions for two B-lines coalesced. (b) The same B-line was detected twice.}
\label{fig:incorrect}
\end{figure}

The pixel-level segmentation networks by construction provide an approximate quantification of the number of B-lines in each frame through counting the isolated origin predictions. The average number of detections per frame, which is currently used to predict B-line presence for a LUS video, can therefore also be considered as a possible measure of B-line severity. In comparison, directly predicting a severity score using a CNN as proposed by Baloescu\etal \cite{baloescu2020automated} is less interpretable and would require retraining and relabeling when there is a need for adjusting the definition of the severity classes. After inspection of our results, we identified several occasions in which further improvement of our method for localization and quantification of B-lines is possible. For example, when two B-line origins are in a close proximity, the predictions can coalesce (e.g., Fig.~\ref{fig:incorrect}a). Although this property is not completely undesirable, as B-lines tend to show similar behavior, it does results in undercounting of B-lines. In addition, we observed that the network can detect multiple origin locations for the same B-line, mostly in cases of pleural consolidation or other bright structures below the pleura (e.g., Fig.~\ref{fig:incorrect}b). This results in counting the same B-line multiple times. Future research can also focus on alternative approaches to quantify the severity of pulmonary congestion in LUS, such as the average intensity value below the pleura \cite{corradi2016computer}, or the fraction of the pleural line from where B-lines originate \cite{brusasco2019quantitative}.

A limitation of this work is that all models were developed and evaluated on LUS videos from a single hospital and acquired using curvilinear transducers from the same vendor. Either a shift in the patient population or acquisition characteristics could cause a mismatch between the data generating distribution that was used for training and a future deployment environment \cite{castro2020causality}. A second limitation is that B-line annotations for training and test data were mainly provided by a single expert. Investigating the generalizability of the developed methods across different clinical sites, as well as extending the curated dataset and acquiring annotations from more LUS experts, are directions for future work.

In conclusion, we compared several deep neural network architectures for detection of B-line artifacts on a new lung ultrasound dataset and proposed a novel approach to B-line localization by predicting the B-line origin locations as single points. 

\bibliographystyle{IEEEtran}
\bibliography{refs.bib}
\end{document}